\documentclass[%
 reprint,
superscriptaddress,
 amsmath,amssymb,
 aps,
prstab,
]{revtex4-1}

\usepackage{graphicx}
\usepackage{dcolumn}
\usepackage{bm}
\usepackage{comment}
\usepackage{caption}
\usepackage{subcaption}

\begin{document}

\preprint{APS/123-QED}

\title{High temporal resolution THz streaking of high brightness relativistic electron beams
}

\author{Maximilian Lenz}
 \affiliation{Department of Physics and Astronomy, UCLA, Los Angeles, CA 90095, USA}
\author{Brian Schaap}%
 \affiliation{Department of Physics and Astronomy, UCLA, Los Angeles, CA 90095, USA}
\author{Atharva Kulkarni}%
 \affiliation{Department of Physics and Astronomy, UCLA, Los Angeles, CA 90095, USA}
\author{Yuemei Tan}
 \affiliation{Department of Engineering Physics, Tsinghua University, Beijing 100084, China}
\author{Yining Yang}%
 \affiliation{Department of Engineering Physics, Tsinghua University, Beijing 100084, China}

\author{Renkai Li}%
 \affiliation{Department of Engineering Physics, Tsinghua University, Beijing 100084, China}
\author{Pietro Musumeci}%
 \affiliation{Department of Physics and Astronomy, UCLA, Los Angeles, CA 90095, USA}

\date{\today}

\begin{abstract}
We report a systematic experimental study of terahertz (THz) streaking structures for ultrafast characterization of relativistic, high-brightness electron beams. Horn-coupled waveguide geometries are investigated, enabling a comparative characterization of streaking strength, dispersion, transmission, and temporal fidelity. Analytical models and electromagnetic simulations are used to describe the dependence of streaking power on the waveguide dimensions and the drive frequency. Experimentally, the structures are characterized using compressed electron beam from an RF photoinjector over a range of THz field strengths, beam energies, and bunch durations. These results establish general design principles and performance limits for THz streaking structures applicable to ultrafast electron beam diagnostics.
\end{abstract}

\maketitle
\section{Introduction}


The generation, control, and measurement of ultrashort, high-brightness electron bunches are central challenges in modern accelerator-based light sources and ultrafast electron beam applications. In particular, femtosecond and sub-femtosecond electron bunches are essential for time-resolved measurements in MeV-scale ultrafast electron diffraction (UED) beamlines \cite{filippetto2022ultrafast, weathersby2015mega, qi2020breaking}, next-generation free-electron lasers (FELs) \cite{emma2010first, duris2020tunable}, inverse Compton scattering sources \cite{hartemann2005high, graves2014compact}, and advanced accelerator concepts requiring precise longitudinal phase-space control \cite{piot2012generation}. In these systems, accurate, single-shot diagnostics of the electron bunch temporal profile and of its relative time-of-arrival with respect to an external laser pulse are critical to maximize performances, guide machine optimization and for quantitative interpretation of experimental data.

Conventional radio-frequency (RF) transverse deflecting cavities have been widely employed for time-resolved beam measurements and direct longitudinal phase-space visualization in many high energy facilities \cite{alesini2006rf, maxson2017direct, akre2008commissioning, behrens2014few}. While RF deflectors provide high streaking strength and excellent temporal resolution \cite{prat2025attosecond}, their size, cost, and limited laser-synchronization capabilities pose challenges for compact beamlines and ultrafast experiments. In particular, the temporal jitter between ultrafast lasers and the high power RF wave that drives these cavities can limit accurate timing determination in pump--probe measurements, motivating alternative approaches that are intrinsically synchronized to the optical clock.

Laser-driven THz streaking has emerged as a powerful alternative for ultrafast electron beam diagnostics \cite{ THz:Streaker:Kealhofer2016, THz:Streaker:li2019terahertz:slit, THz:Streaker:zhao2018terahertz:slit}. In this technique, an intense THz pulse imparts a time-dependent transverse momentum to an electron bunch, mapping the longitudinal charge distribution onto a transverse spatial profile after a drift. Because the THz field is derived from the same laser system that drives photoemission or optical pumping, this technique enables single-shot time-of-arrival and bunch length measurements with intrinsic femtosecond-level synchronization. These features have made THz streaking particularly attractive for few MeV UED beamlines \cite{THz:Streaker:othman2023measurement:Streak_Compress, THz:Compr:snively2020femtosecond} and compact accelerator platforms where laser-driven THz waves are already available \cite{THzAccel:nanni2015terahertz,THzAccel:zhang2018segmented,THz:Accel:hibberd2020acceleration:waveguide,THz:accel:zhang2020cascaded,THz:accel:xu2021cascaded,THz:accel:tang2021stable}.

Naturally, one might wonder about the scalability of THz streaking to electron beamlines with higher beam energies (10-1000 MeV). There are two main factors to take into account. On one hand, the angular kick imparted by a THz field scales approximately as $1/\gamma$, where $\gamma$ is the Lorentz factor of the electron beam. On the other hand, for a given spot size, the intrinsic angular divergence of a high-brightness electron beam, set by its geometric emittance, also scales inversely with $\gamma$. In that sense, the streaking-induced modulation does not necessarily vanish in relative terms. In practice, however, the reduced deflection in absolute terms tightens the requirements on the design of the transport after the THz interaction point and on the detector screen resolution. More importantly, there is a need for higher streaking strength to be truly competitive with RF transverse deflectors, motivating a significant (ideally up to an order-of-magnitude) improvement in coupling and field enhancement for compact THz structures.

Significant efforts have been devoted to increasing the achievable streaking strength of THz-based diagnostics through field enhancement, dispersion control and lengthening of the interaction. Approaches include split ring resonators \cite{fabianska2014split}, velocity matching in dielectric materials \cite{THz:Streaker:volkov2022spatiotemporal}, crossing interaction geometries \cite{THz:Streaker:baek2021real:90degstreaking}, and horn-coupled metallic waveguides \cite{yang2025sub}. Among these, horn-coupled waveguide geometries provide a promising combination of efficient THz coupling, strong field confinement, and mechanical robustness. However, the performance of such structures depends critically on the geometric parameters, the input beam energy and the characteristics of the driving THz pulse, so that a systematic experimental comparison of different designs remains limited.

In this work, we present a systematic experimental study of horn-coupled THz waveguide structures for ultrafast electron beam diagnostics carried out at the UCLA Pegasus photoinjector laboratory \cite{alesini2015new}. Different waveguide geometries are investigated and compared in terms of streaking strength, dispersion, transmission efficiency, and temporal fidelity. Analytical models and electromagnetic simulations are used to elucidate the dependence of performance on the structure length and its transverse dimensions, and operating frequency. The structures are experimentally characterized using the RF-compressed relativistic electron beam from the UCLA Pegasus photoinjector \cite{maxson2017direct}, with calibrated deflection measurements performed over a range of THz field strengths, beam energies, and bunch durations. These measurements validate our models and define practical design rules and performance limits for compact THz streakers, thus enabling us to scale the results across a wide range of electron beam energies. 

\section{Waveguide structure theory}
\label{sec:theory}
\begin{figure}[htbp]
    \includegraphics[width=0.9\columnwidth]{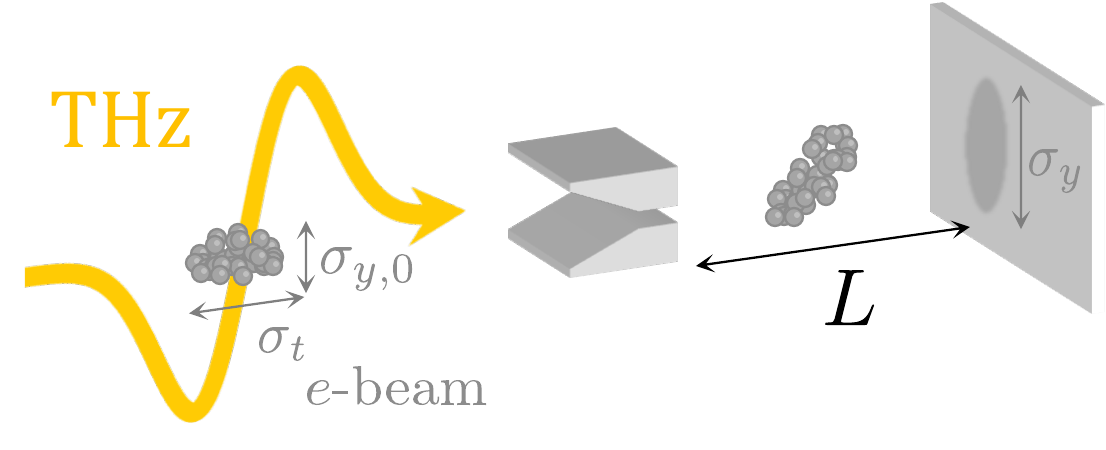}
    \caption{Schematic of horn-coupled hollow waveguide THz streaking. The beam arrives at the structure at the steepest  zero crossing in the THz wave. Electrons at the head and tail of the beam sample opposite THz phases and receive opposite transverse kicks, mapping longitudinal position within the bunch onto transverse position in the streaking plane on a downstream screen.}
    \label{fig:cartoon}
\end{figure}

The experimentally observable figure of merit for a transverse deflecting structure is the shearing strength parameter, $S$, defined as the rate of change of the transverse deflection angle with respect to the longitudinal time coordinate of the beam (see Fig. \ref{fig:cartoon} for a schematic of the interaction). For a relativistic electron beam traversing a deflecting field, we can write:
\begin{equation}
    S = \frac{d(\Delta y')}{dt} \approx \frac{\omega e V_{\perp, \text{eff}}}{E_b}
    \label{eq:shear_strength}
\end{equation}
where $\omega$ is the angular frequency of the streaking field, $E_b$ is the beam energy, and $V_{\perp, {\rm eff}}$ is the effective transverse deflecting kick integrated along the interaction length. This expression assumes that the electrons sample the locally linear field near the zero crossing of the sinusoidal deflecting waveform.

Let us consider a hollow rectangular waveguide operating in the fundamental TE$_{10}$ mode as the streaking structure. Unlike dielectric-lined structures that slow the phase velocity $v_{ph}$ to match the beam velocity ($v_b \approx c$) \cite{THz:Streaker:lemery2017transverse:DLW,THz:Streaker:georgiadis2021dispersion:DLW,THz:Structure:lake2022longitudinal:DLW}, we operate in the ``fast wave'' regime where $v_{ph} \geq c$. The transverse Lorentz force experienced by an electron with normalized velocity $\beta = v_b/c$ is written as
\begin{equation}
    F_y = e(E_y - v_b B_x) = e E_y \left( 1 - \frac{\beta}{\beta_{ph}} \right)
    \label{eq:lorentz_force}
\end{equation}
where the net deflection arises from the incomplete cancellation between the electric and magnetic forces and $B_x = E_y/\beta_{ph}c$ in the waveguide. 

To achieve a non-zero kick, the interaction must be terminated before the electron slips into the opposite sign phase of the wave. The optimal interaction length is set by the dephasing length, $L_d$, defined as the distance over which the electron slips by $\pi$ radians relative to the wave phase:
\begin{equation}
    L_d = \frac{\pi}{k_0 \left| \frac{1}{\beta} - \frac{1}{\beta_{ph}} \right|}
    \label{eq:dephasing_length}
\end{equation}
where $k_0 = \omega/c$. The integrated effective voltage over a structure of length $L$ can then be written as
\begin{equation}
    V_{\perp, \text{eff}} = E_\mathrm{max} L \left( 1 - \frac{\beta}{\beta_{ph}} \right) \text{sinc}\left( \frac{\pi L}{2 L_d} \right) \eta_{\text{loss}}
    \label{eq:effective_voltage}
\end{equation}
where $E_\mathrm{max}$ is the peak transverse electric field on axis and $\eta_{\text{loss}}$ accounts for attenuation losses.

The waveguide dimensions ($a \times b$) can be optimized to balance field intensity, synchronization, and attenuation. The width $a$ determines the cutoff frequency $f_c = c/2a$ and the dispersive properties of the waveguide. The phase velocity of the deflecting mode $\beta_{ph} = [1 - (f_c/f)^2]^{-1/2}$, for $a = 750~\mu\mathrm{m}$ excited by a THz pulse with center frequency of 0.5~THz, can be calculated as $\beta_{ph} \approx 1.09$, yielding a dephasing length longer than 3.5~mm for ultrarelativistic beams.

The vertical slit height $b$ acts as a field concentrator. For a fixed input power $P$, the field scales as $E_y \propto b^{-1/2}$. However, the attenuation coefficient due to wall losses is inversely proportional to $b$ for $a \gg b$ \cite{zangwill}. Our choice of $b = 50~\mu\mathrm{m}$ maximizes the local gradient, but it introduces significant ohmic losses, estimated at $\approx 1.5\,\mathrm{dB/cm}$ for copper with realistic surface roughness (assuming $2\times$ theoretical losses). In order to couple free-space THz radiation into the sub-wavelength gap, we employ a tapered horn with a profile optimized in order to minimize mode conversion losses \cite{stutzman2012antenna}.

\begin{table}[ht]
    \caption{\label{tab:parameters}Parameters for Streaking Estimates}
    \begin{ruledtabular}
        \begin{tabular}{lcc}
            Parameter & Symbol & \multicolumn{1}{c}{Value} \\
            \colrule
            Kinetic Energy & $E_\mathrm{b}$ & 4.6~\text{MeV} \\
            THz Energy & $U_\mathrm{THz}$ & 20~$\mu$\text{J} \\
            Center Frequency & $f_0$ & 0.5~\text{THz} \\
            Rectangular Waveguide Length & $L$ & 2.4~\text{mm} \\
            Aperture & $a \times b$ & 750 $\times$ 50\,$\mu$\text{m} \\
            PPWG length & L$_\mathrm{ppwg}$ & 8 \,\text{mm}\\
            PPWG height & h & 100\,$\mu$\text{m} \\
        \end{tabular}
    \end{ruledtabular}
\end{table}

Based on these considerations, we can estimate the expected streaking performance for the beam and THz parameters used in our experiments, summarized in Table~\ref{tab:parameters}. The input THz waveform is assumed to be a single-cycle pulse centered at 0.5~THz with an energy of 20~$\mu$J. Accounting for a conservative 50\% coupling efficiency, including Fresnel reflections and mode mismatch, approximately 10~$\mu$J is coupled into the waveguide. 

\begin{figure}[htbp]
    \centering
    \includegraphics[width=0.9\columnwidth]{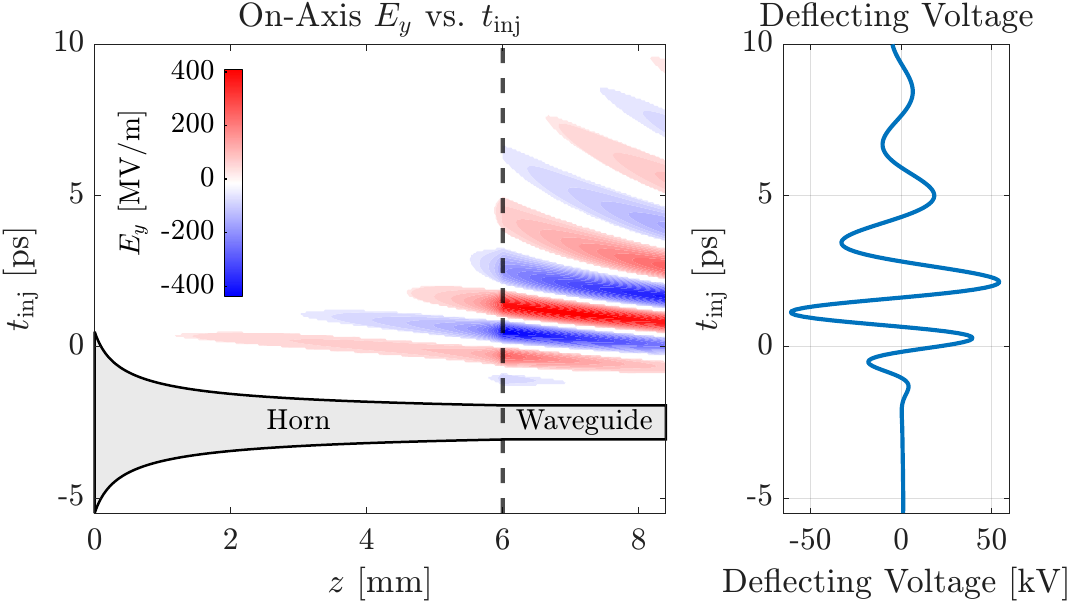}
    \caption{a) Map of the field experienced along the structure (longitudinal coordinate $z$) by electrons injected at time $t$ when fed by a single-cycle THz pulse waveform centered at 0.5 THz. b) Transverse deflecting voltage experienced by a 4.6\,MeV electron as a function of the relative time of injection into the structure.}
    \label{fig:fieldmap}
\end{figure}

An electromagnetic simulation of the single-cycle THz waveform evolution in the waveguide structure is shown in Fig.~\ref{fig:fieldmap}. The simulation shows that significant dispersion occurs during propagation in the guide: the initially single-cycle waveform stretches to more than 1.5~ps FWHM. This temporal broadening reduces the peak power of the coupled THz pulse to approximately $P \simeq 6$~MW. The peak electric field $E_\mathrm{max}$ inside the waveguide can then be estimated from the power-flow relation for the TE$_{10}$ mode:
\begin{equation}
    E_\mathrm{max} = \sqrt{\frac{4 P Z_\mathrm{TE}}{a b}} \approx 375~\mathrm{MV/m},
    \label{eq:max_field}
\end{equation}
where $Z_\mathrm{TE} = Z_0/\sqrt{1-f_c^2/f^2} \approx 411~\Omega$ is the wave impedance at 0.5~THz, with $Z_0 = 377~\Omega$ the free-space impedance.

For electrons with kinetic energy 4.6~MeV and a central THz drive frequency of 0.5~THz, the magnetic cancellation factor is $(1-\beta/\beta_\mathrm{ph}) = 0.09$. The sinc-like slippage factor over the final 2.4~mm interaction length is approximately 0.81. Including attenuation losses, $\eta_\mathrm{loss} \approx 0.94$, the corresponding effective transverse voltage is
$V_\perp \simeq 60$~kV, yielding a projected shearing strength of 
$S \approx 37~\mu\mathrm{rad/fs}$. Depending on the intrinsic beam divergence, this streaking strength is sufficient to enable sub-fs temporal resolution in bunch-profile measurements.

The full trajectory-integrated deflecting voltage obtained from the electromagnetic simulation is shown in Fig.~\ref{fig:fieldmap}(b) and is in good agreement with the estimate above. Two features stand out. First, the net streaking kick is generated predominantly in the final $\simeq 2.4$\,mm of the straight waveguide, while the horn contributes only weakly. This is not surprising: the horn primarily acts as an impedance transformer and mode launcher, and the field there is less confined and less velocity-matched to the on-axis electrons, so its contribution to the trajectory-integrated transverse Lorentz force is small compared with the region where the guided mode is fully established and concentrated near the beam axis. Second, the tilted field contours directly reveal the slippage between the THz pulse and the relativistic electrons. Because the phase velocity in the waveguide is superluminal ($v_\mathrm{ph}>c$), the field contours are tilted downward in the picture, i.e. the zero crossing overtakes the electrons as they propagate in the guide. At the same time, since the group velocity in the waveguide is subluminal ($v_g<c$), the electrons gradually advance with respect to the pulse envelope as they propagate, which appears to lag behind (i.e. towards positive time) as the wave propagates. As a result, optimal streaking is obtained when the e-beam is injected slightly after the THz pulse at the entrance (positive times in the figure), so that the electrons remain close to the peak deflecting phase over the interaction length.

\begin{figure}[htbp]
        \includegraphics[width=0.95\linewidth]{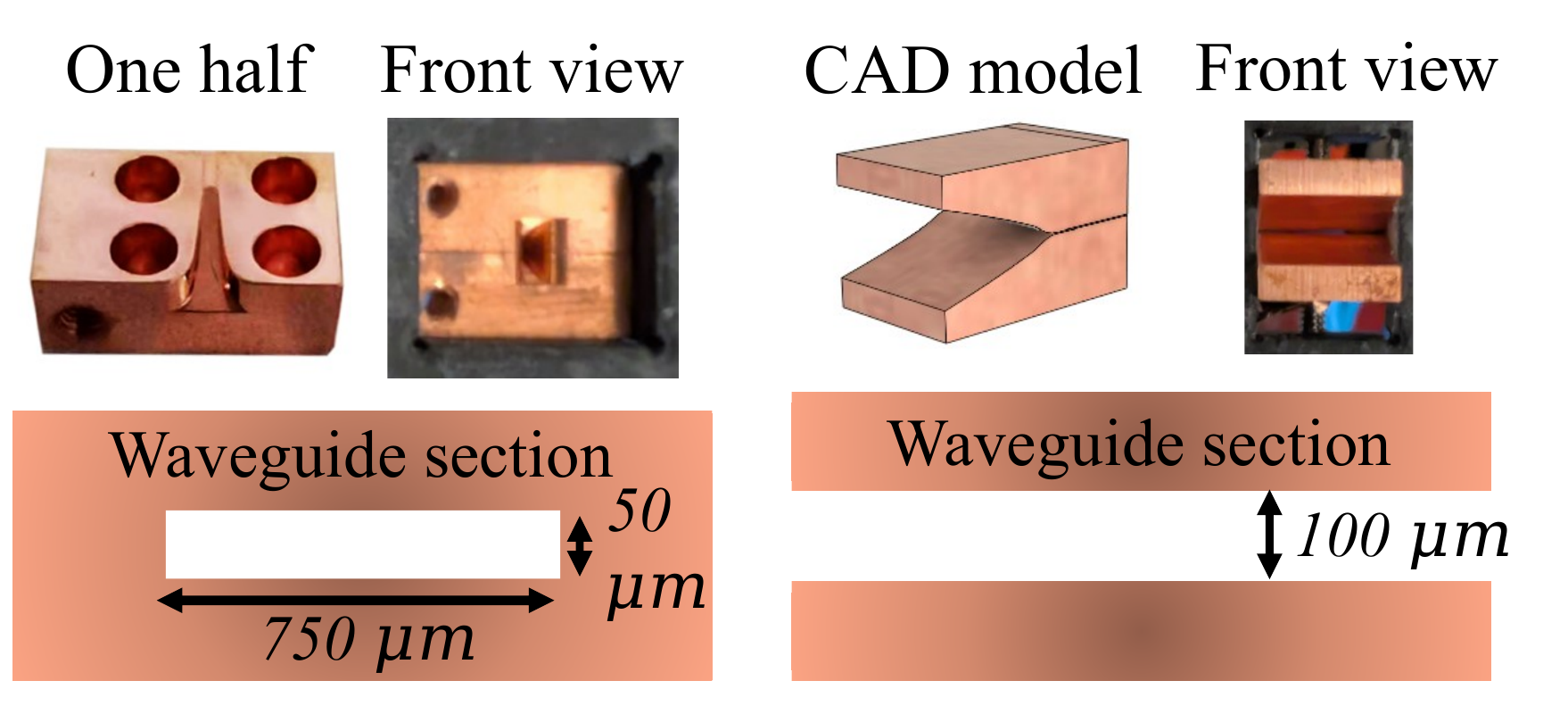}
    \caption{THz structures characterized at UCLA Pegasus laboratory with waveguide dimensions illustrated: a) Rectangular waveguide with horn designed at Tsinghua university. b) Parallel plate waveguide with tapered aperture designed at UCLA.}
    \label{fig:THzstructures}
\end{figure}

In contrast to the closed rectangular geometry, a parallel plate waveguide (PPWG) supports a fundamental transverse electromagnetic (TEM) mode. This mode exhibits no low-frequency cutoff and propagates with a phase velocity identically equal to the speed of light ($v_{ph} = c$). The primary advantage of the TEM mode is the absence of group velocity  dispersion, which preserves the temporal profile of broadband, single-cycle THz pulses over long propagation distances. Furthermore, for relativistic beams ($\beta \approx 1$), the $v_{ph} = c$ condition significantly mitigates the longitudinal dephasing that restricts the interaction length in fast-wave structures. For a $\gamma = 10$ electron beam, the dephasing length $L_d = \lambda / [2(1/\beta - 1)]$ extends to several centimeters. Instead of dephasing, the effective interaction length in a PPWG is inherently limited by transverse diffraction. Because the TEM mode is unconfined in the horizontal plane, the THz wave propagates as a 1D Gaussian beam. The interaction distance is bounded by the Rayleigh range, $z_R = \pi w_x^2 / \lambda$, where $w_x$ is the horizontal beam waist. Consequently, the structure length is optimized to match the confocal parameter ($L \approx 2 z_R$), ensuring the electron bunch interacts primarily with the high-intensity focal region before diffractive spreading significantly diminishes the field amplitude. 

To estimate the peak on-axis electric field $E_0$, assuming 35\% reflection loss from the coupling taper, more than $16\,\mathrm{MW}$ of peak power is delivered to the interaction region. The power is significantly higher as in this case the pulse can propagate without dispersion. For a PPWG with a vertical gap of $b = 100\,\mu\mathrm{m}$ and a tightly focused horizontal waist of $w_x \approx 400\,\mu\mathrm{m}$, the peak electric field can be calculated from the cycle-averaged TEM intensity relationship, $E_0 = \sqrt{2 Z_0 P / (w_x b)}$, yielding a localized peak field of $E_0 \approx 550\,\mathrm{MV/m}$.

However, the transverse streaking of relativistic electron beams in this geometry faces a fundamental limitation due to magnetic cancellation. In an ideal plane-wave TEM mode, the orthogonal magnetic field is precisely $B_x = E_y/c$, leading to a $(1-\beta)$ suppressed Lorentz force. This cancellation is mitigated by the fact that in the focused geometry the wave is not a perfect plane wave. The Gouy phase shift, $\zeta(z) = \frac{1}{2} \arctan(z/z_R)$ introduces a spatially varying phase which modifies the local effective wavenumber, $k_{\text{eff}} = k_0 - \partial_z \zeta$, rendering the phase velocity locally superluminal near the focus and partially lifting the magnetic suppression.

Assuming a structure length of $L = 8\,\mathrm{mm}$, the effective deflecting voltage in a bare metal PPWG can be integrated to $V_{\perp, \text{eff}} \approx E_0 L (1-\beta) \approx 24\,\mathrm{kV}$ which is smaller than what achieved with the shorter fast-wave rectangular waveguide. The main benefit of this configuration is that the pulse does not stretch in time and preserves the single-cycle nature. The fabrication process is also simplified as two halves are separately machined and then assembled together at the desired gap spacing. 

\section{Experimental Setup}

\subsection{Electron beamline}

The streaking performance of the THz deflecting structures was characterized experimentally on the UCLA Pegasus relativistic electron beamline \cite{alesini2015new}. The drive laser is a Ti:sapphire chirped-pulse-amplification system delivering up to 20\,mJ pulses at 780\,nm, with 100\,fs duration at a repetition rate of 10\,Hz. Approximately 1\,mJ of the infrared output is diverted for third-harmonic generation to produce ultraviolet (UV) pulses for photoemission, while the remaining energy is used for laser-driven THz generation and electro-optic sampling (EOS)-based characterization.
 
\begin{figure*}[htb]
    \includegraphics[width = 0.8\textwidth]{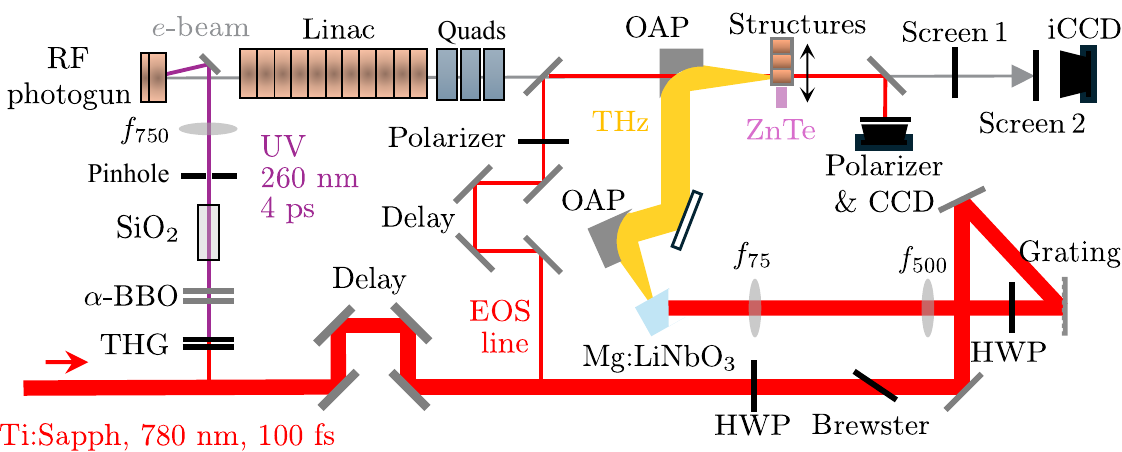}
    \caption{Schematic of the beamline layout and THz source implementation. The Ti:sapphire laser is split into three arms: one is frequency-tripled and temporally shaped to drive photoemission, a second is used to generate high-power THz pulses by optical rectification for the streaking structures, and a third provides the probe for electro-optic diagnostics used in alignment and field calibration.}
    \label{fig:Experimental Setup}
\end{figure*}

As shown in Fig.~\ref{fig:Experimental Setup}, photoelectrons are generated from a Cs$_2$Te photocathode using 260\,nm UV pulses obtained by third-harmonic conversion of the Ti:sapphire laser. To control the temporal profile of the emitted electron bunch, the UV pulses can be optionally shaped using two $\alpha$-BBO crystals with thicknesses of 1.09~mm, and 2.18~mm oriented at $45^\circ$ with respect to one another and a 20~cm long fused-silica dispersive rod to form an approximately flat-top temporal profile with a total duration of $\sim 4$~ps FWHM.

The UV beam is then transmitted through a 100\,$\mu$m pinhole and imaged onto the photocathode of a 1.6-cell S-band RF gun, producing a round laser spot with rms radius below 10\,$\mu$m. The gun generates electron bunches with 3.2\,MeV kinetic energy, bunch charge tunable from 1 to 300\,fC through the incident laser energy, and normalized emittance below 20\,nm-rad~\cite{maxson2017direct}. Downstream of the gun, an 11-cell S-band linac can accelerate the beam up to 7.5\,MeV or operate off-crest to impose a longitudinal phase-space chirp for bunch compression. The beam energy is measured with a dipole spectrometer. Before entering the THz streaking structure, the beam is focused by a quadrupole triplet located 1.7\,m downstream of the linac. The quadrupoles are tuned to maximize transmission through the THz structure aperture at a plane 0.55\,m downstream of the triplet, while maintaining a small vertical beam size at the observation screens located at 0.42\,m and 1.27\,m respectively from the interaction. Under typical operating conditions, transmission through the $50\times750\,\mu\mathrm{m}^2$ rectangular slit is below 5\%, and the vertical rms beam size at the main downstream screen is approximately $\sigma_y \approx 80\,\mu$m. 

\subsection{THz source}

High-field THz pulses are generated via optical rectification in stoichiometric Mg-doped LiNbO$_3$ (sLN), which is well suited for high-energy THz generation due to its large nonlinear optical coefficient, high optical damage threshold, and scalability to centimeter-scale crystals \cite{THz:Generation:wu2023generation:record}. Efficient THz generation in LiNbO$_3$ requires compensation of the large velocity mismatch between the near-infrared pump pulse ($n_{\mathrm{gr}}^{\mathrm{IR}} \approx 2$) and the generated THz radiation ($n^{\mathrm{THz}} \approx 5$), which is achieved by pulse-front-tilting (PFT) the pump pulse \cite{Generation:kroh2022parameter,Generation:Hebling:theory,THz:hebling2002velocity:Generation}.

For the experiments discussed here, the optimum PFT angle was determined to be $\gamma_{\mathrm{PFT}} = 62.6^\circ$, from the material dispersion of stoichiometric LiNbO$_3$ at room temperature \cite{THzcrystal:gayer2008:sLN-IR,THz:LN:palfalvi2005temperature:THz-refractive-index,THz:Generation:nakamura2002optical:refractive-index-sLN,THzcrystal:buzady2020:sLN-THz}. The sLN crystal is a prism with a $9 \times 9$\,mm$^2$ pump entrance surface and an exit face cut at $62^\circ$. PFT is generated using an 830\,lines/mm diffraction grating with a $30 \times 30$\,mm$^2$ square aperture. To minimize pulse-front distortions and spatial chirp, the first diffraction order is selected at $\theta_d = 0^\circ$ \cite{Optics:kreier2012:PFTdistortions}. While this arrangement decreases the grating reflection efficiency, the laser provides enough energy that after aperture clipping and transport losses, up to 6\,mJ are available for pumping of the sLN crystal.

The grating is imaged into the sLN using a two-lens telescope configuration, which allows independent control of the imaging plane and magnification. Compared to a single-lens imaging scheme, this approach provides increased robustness against longitudinal misalignment and maintains a nearly constant pulse-front tilt over an extended interaction region inside the crystal \cite{THz:Generation:lenz2023multipulse,THz:Generation:tokodi2017optimization:PFT-optimization}. The telescope consists of lenses with focal lengths of 500\,mm and 75\,mm arranged in a near-Kepler configuration, resulting in a nominal magnification of $M \approx 0.15$. Half-wave plates placed before and after the grating are used to optimize diffraction efficiency and to reorient the pump polarization for efficient THz generation.

To improve THz outcoupling efficiency, a thin Kapton film (Thorlabs KAP22-075) is applied to the output surface of the crystal to reduce Fresnel losses at the crystal-air interface, resulting in an approximately 20\% increase in extracted THz energy. The THz pulse energy is measured using a calibrated pyroelectric detector showing shot-to-shot energy stability of approximately 5\%. As shown in the next section of the paper, at a pump energy of 6\,mJ incident on the crystal, an IR-to-THz conversion efficiency of up to 0.5\% is measured, corresponding to a THz pulse energy of approximately 30\,$\mu$J. This conversion efficiency at room temperature, to the best of our knowledge, has only been exceeded with a stair-step-echelon PFT scheme \cite{THz:Generation:guiramand2022near:high_Efficiency}.

The generated THz radiation is collimated using an off-axis parabolic (OAP) mirror, redirected by a flat mirror, and then focused into the streaking structure by the second OAP containing a 3\,mm diameter central aperture to allow co-propagation of the electron beam (see Fig.~\ref{fig:Experimental Setup}). Precise alignment of the OAPs is critical, as small transverse or angular misalignments lead to significant aberrations at the focus. Because the crystal exit surface is cut at $62^\circ$, slightly different from the optimal pulse-front tilt, the generated THz beam exits the crystal at a small angle relative to the surface normal. After refraction at the crystal-air interface, this results in an external propagation angle of approximately $3^\circ$ that is crucial to take into account for positioning the first OAP \cite{THz:Generation:gabriel2025three:near-field}. Alignment is performed by back-propagating visible light from the focus of the second OAP to the nonlinear crystal. An EOS line is implemented by splitting off less than $10\%$ of the pump energy using a pellicle beamsplitter. A crossed-polarizer configuration is implemented, providing a simple method to verify spatial overlap and confirm location and dimensions of the THz focus \cite{THz:Generation:lenz2023multipulse}. For quantitative measurements, the second polarizer is replaced by a quarter-wave plate and a Wollaston prism, enabling balanced detection enabling full temporal and spectral characterization of the THz waveform at the focus of the OAP. After propagation through air and an HDPE vacuum window, approximately 38\% of the THz energy is lost, resulting in a pulse energy just below $\sim 20\,\mu$J at the focus of the streaking structure.

\section{Streaking Calibration}

\begin{figure}[ht]
\centering
\includegraphics[width=1\columnwidth]{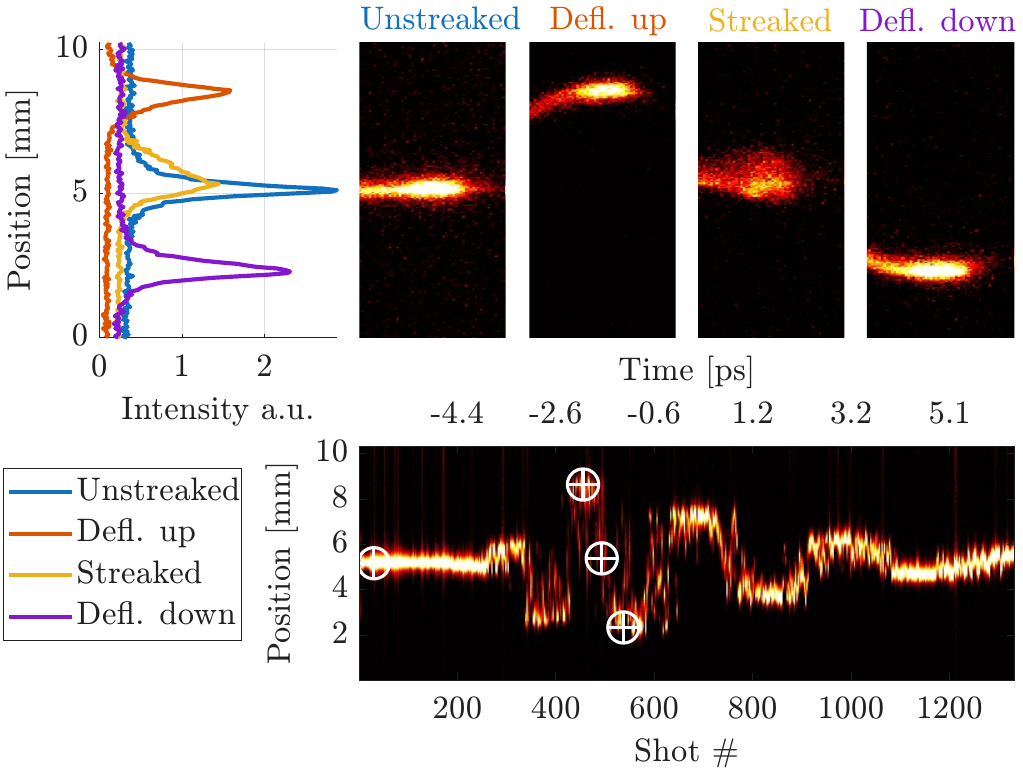}
\caption{The top right row depicts four raw images of electron beams with different relative time of arrival with respect to the THz wave captured on the first screen 42\,cm downstream. The individual y-projections on the left are color coded in reference to the titles of each image. The bottom plot shows a heatmap of the projection for each individual image while during the temporal delay is adjusted. The highlighted points show the temporal positions corresponding to the four shots displayed in the top.}
\label{fig:streaking_raw}
\end{figure}

A key figure of merit for the deflecting structures is the observable shearing strength, defined as the time derivative of the transverse angular deflection, $d\theta/dt$, conveniently expressed in units of $\mu$rad/fs. In practice we obtain a deflection waveform by scanning the relative delay of the THz pump pulse with respect to the electron beam and recording the resulting beam distribution on a screen located 42\,cm downstream of the streaking structure. For each delay setting, approximately 15 shots of the transverse beam distribution are recorded. Representative raw single shot images obtained with the rectangular waveguide structure at the compressing linac phase are shown in Fig.~\ref{fig:streaking_raw}. The observed transverse curvature of the deflected beam on the screen is an effect of TE$_{10}$ mode spatial field profile within the waveguide. In order to limit the analysis to particles that pass through the center of the structure and experience the largest field gradient, we restrict the region of interest to the central area of the beam.

In practice, two effects complicate the reconstruction of the THz deflection waveform from a delay scan: (i) if the electron bunch spans a large temporal window within the THz cycle, a significant distortion of the projected transverse distribution can occur requiring some care to estimate the THz-induced deflection, and (ii) any shot-to-shot arrival-time jitter smears the deflection-versus-delay trace and makes a simple average at each delay an unreliable proxy for the underlying waveform. 

\begin{figure*}[htbp]
        \includegraphics[width=0.46\linewidth]{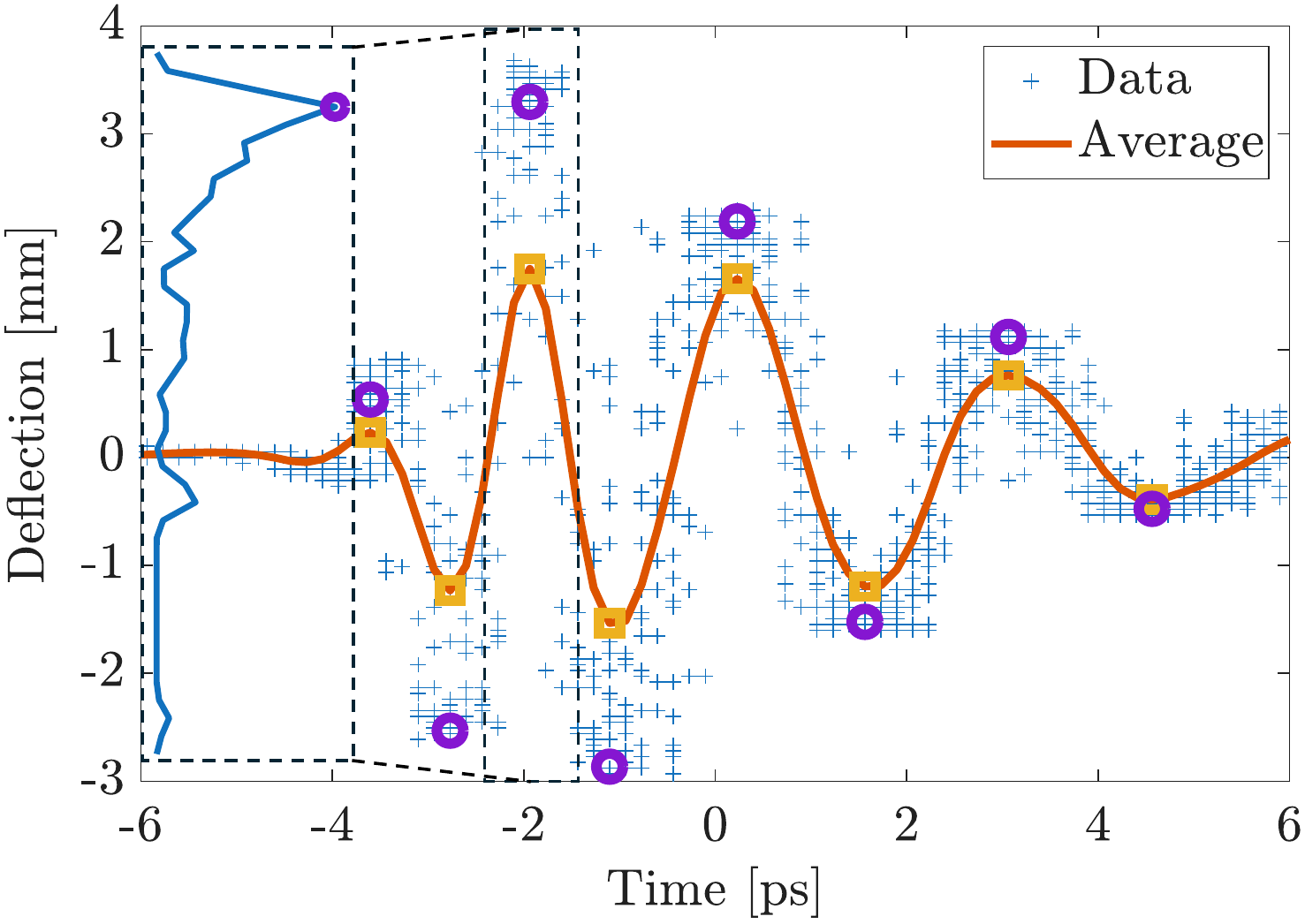}
        \includegraphics[width=0.5\linewidth]{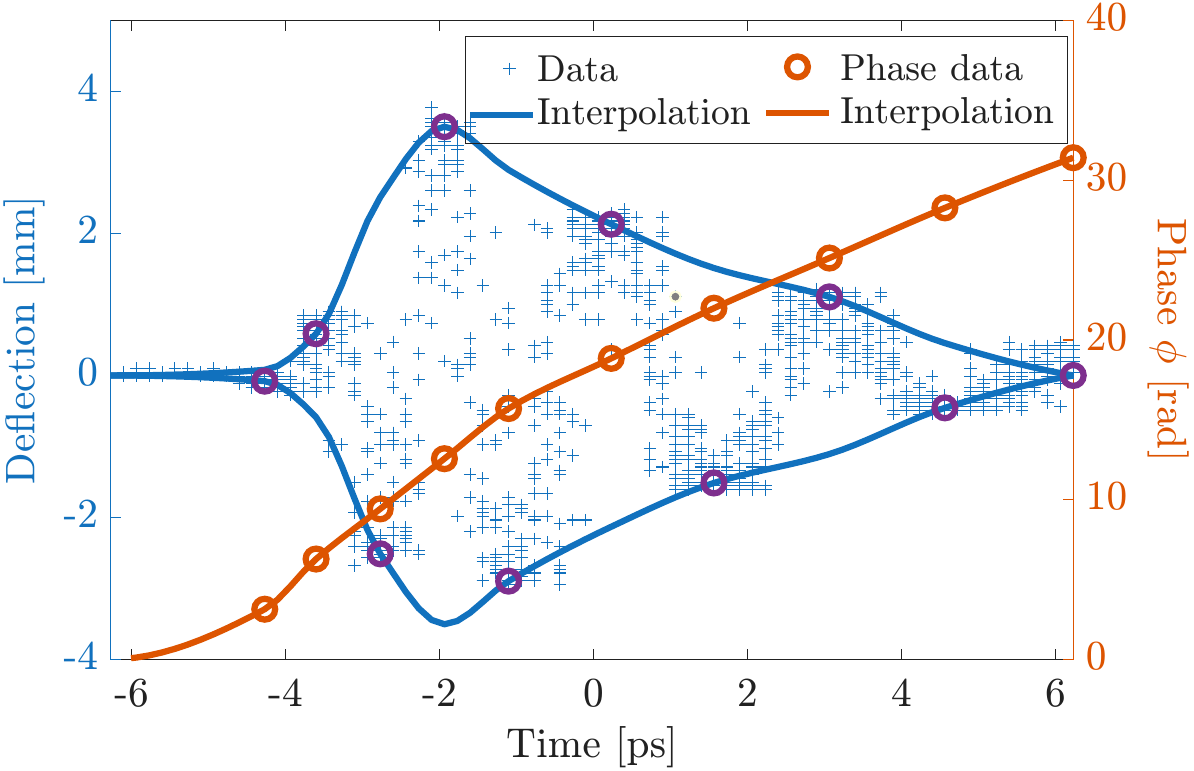}
    \caption{Extrema-based waveform reconstruction procedure: a) Identification of extrema via windowed accumulation. b) Amplitude and phase interpolation used for waveform reconstruction.}
    \label{fig:Waveform_reconstruction}
\end{figure*}

The first step is to determine the deflection of an individual electron bunch. If the electron bunch is sufficiently short compared to the THz wavelength, all electrons probe nearly the same phase of the streaking wave and the distribution centroid accurately reflects the induced deflection. However, for bunches that span more than 10 degrees of phase of the wave, larger time windows in the deflection waveform are sampled simultaneously, producing asymmetric tails in the screen projection as clearly seen in the projection in Fig.~\ref{fig:Waveform_reconstruction}. For this reason, we used the peak or \emph{mode} of the projected transverse distribution rather than its centroid to retrieve the maximum transverse displacement and then the angular deflection after dividing by the structure-screen propagation distance.


The second step is reconstruction of the deflection waveform itself. In practice, extraction of the waveform is complicated by significant shot-to-shot arrival-time jitter of the electron beam. At the UCLA Pegasus laboratory, this jitter is on the order of $\sim 300$\,fs. This jitter arises from RF-to-laser phase fluctuations at the photocathode and beam energy variations affecting the time of flight. Because this jitter is comparable to half cycle of the highest-frequency components of the THz pulse, conventional deconvolution or phase-retrieval techniques are not reliable.

To overcome this limitation, we employ an extrema-based reconstruction technique that is robust against timing jitter, leveraging the fact that the local slope of the deflection is small near maxima/minima and so, at these points timing jitter maps to relatively little variation in the data. The procedure to identify the extrema can be followed in Fig.~\ref{fig:Waveform_reconstruction}. Practically, we (1) compute the modal deflection shot-by-shot and form a lightly smoothed average versus delay to locate approximate extrema times; (2) around each extremum, collect all shots within a $\sim$300\,fs delay window; (3) pool these shots and take the mode of the accumulated deflection distribution as the extremal deflection.
 
With the extrema identified, the deflection waveform is reconstructed assuming a smooth form
\begin{equation} 
y(t) = A(t)\cos[\phi(t)],
\end{equation}
where $A(t)$ is the amplitude envelope and $\phi(t)$ the instantaneous phase. The amplitude is obtained by interpolating the extrema, while enforcing the phase $\phi(t_{\mathrm{ext}}) = n\pi$ at successive extrema. The resulting amplitude and phase functions for a representative dataset are shown in Fig.~\ref{fig:Waveform_reconstruction}. Although this approach cannot retrieve sub-cycle phase information, it provides a robust representation of the waveform sufficient for determining the streaking strength even for long bunches and in the presence of substantial arrival-time jitter. This quantity is finally obtained by converting the reconstructed deflection to angular units and evaluating $d\theta/dt$, with particular emphasis on the maximum slope near the zero crossings. 

\begin{figure*}[htbp]
        \includegraphics[width=0.48\linewidth]{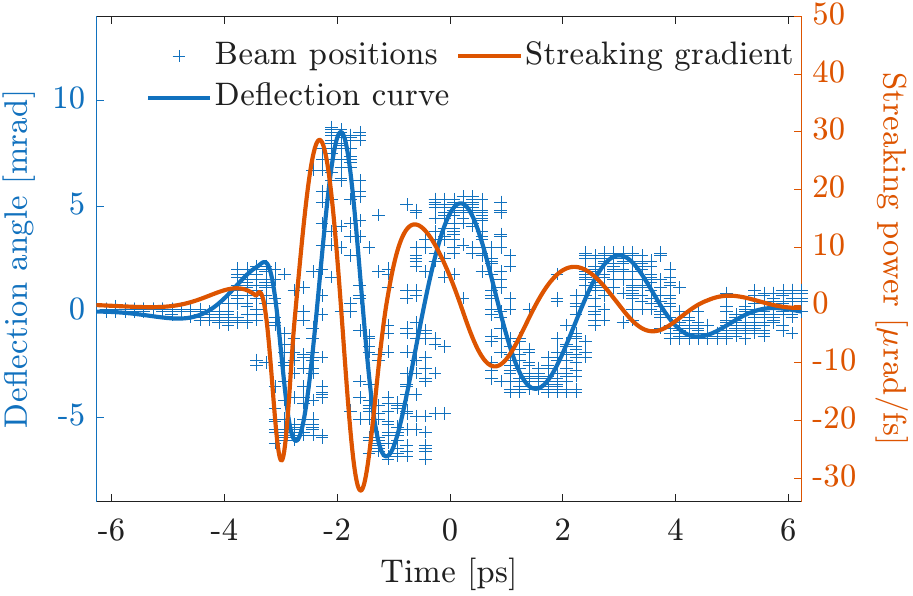}
        \includegraphics[width=0.48\linewidth]{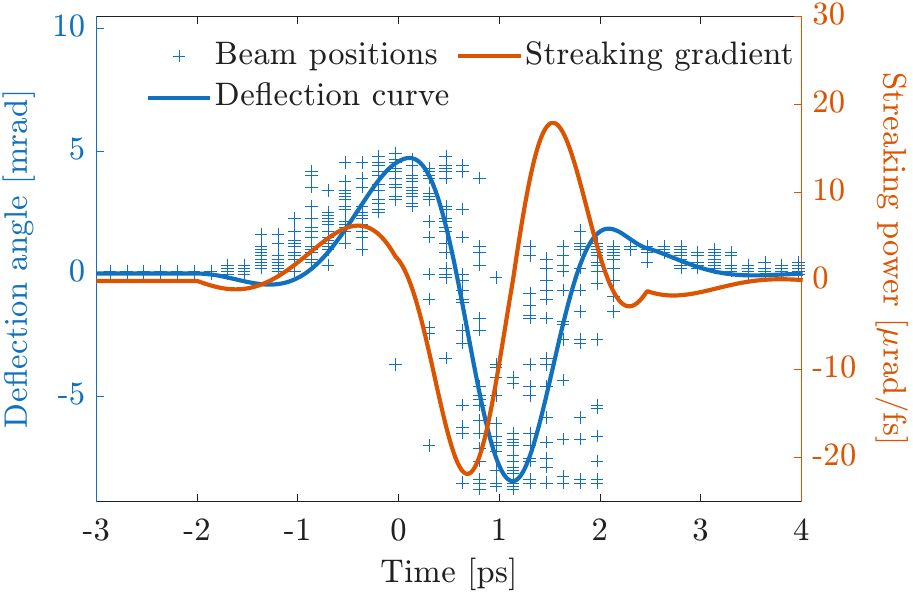}
    \caption{Reconstructed waveform for the horn-coupled rectangular waveguide (a) and parallel plate waveguide (b) described in section \ref{sec:theory}.}
    \label{fig:Waveforms}
\end{figure*}

The reconstructed waveforms for the horn-coupled rectangular waveguide and the parallel-plate waveguide are shown in Fig.~\ref{fig:Waveforms}. While both structures produce comparable peak deflection angles of approximately 8~mrad, the horn-coupled rectangular waveguide exhibits a steeper slope due to enhanced high-frequency components supported by the closed waveguide geometry. The rectangular waveguide achieves a peak streaking gradient of $-(32\pm2)~\mu$rad/fs, compared to $-(21\pm2)~\mu$rad/fs for the parallel-plate waveguide, demonstrating the importance of spectral confinement and mode control in maximizing temporal shear.

With the reconstructed deflection waveform, the local deflection gradient $\mathrm{d}y/\mathrm{d}t$ is known at each temporal position, enabling an estimate of the timing jitter. The shot-to-shot transverse position jitter on the screen, denoted $\sigma_y$, can be converted into an equivalent temporal jitter via $\sigma_t = \sigma_y / |\mathrm{d}y/\mathrm{d}t|$. Evaluating this quantity across the delay scan and averaging over all time delays yields a characteristic temporal jitter of $\sigma_t \approx 250\,\mathrm{fs}$, consistent with independent estimates of arrival-time jitter at the Pegasus beamline further confirming the robustness of the extrema-based reconstruction method.

\section{Structure Characterization} 

The performance of the THz deflecting structures is then experimentally characterized as a function of the incident THz pulse energy and varying the electron beam energy. The first measurement quantifies how efficiently electromagnetic energy delivered to the structure is converted into transverse deflecting voltage, while the second directly validates the inverse beam-energy scaling predicted by Eq.~\eqref{eq:shear_strength}. Together, these datasets establish both the efficiency and the achievable temporal shear of the structures and enable extrapolation of the performance of these structures at higher energy and a quantitative comparison to conventional RF-based transverse deflecting cavities. Owing to its superior performance, the rectangular waveguide horn geometry is emphasized in this section.

\begin{figure}[ht]
\centering
\setlength{\unitlength}{1cm}
\begin{picture}(10,7)
  \put(0,0){\includegraphics[width=1\columnwidth]{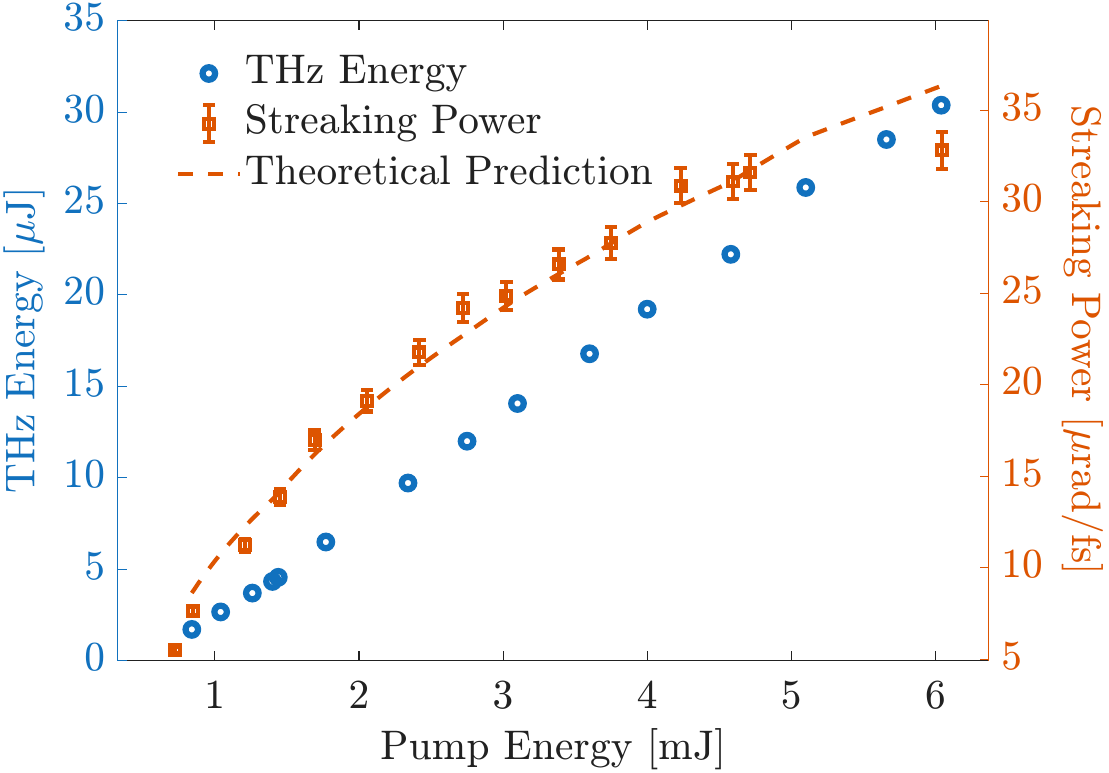}}
  \put(4.7,0.9){\includegraphics[width=0.32\columnwidth]{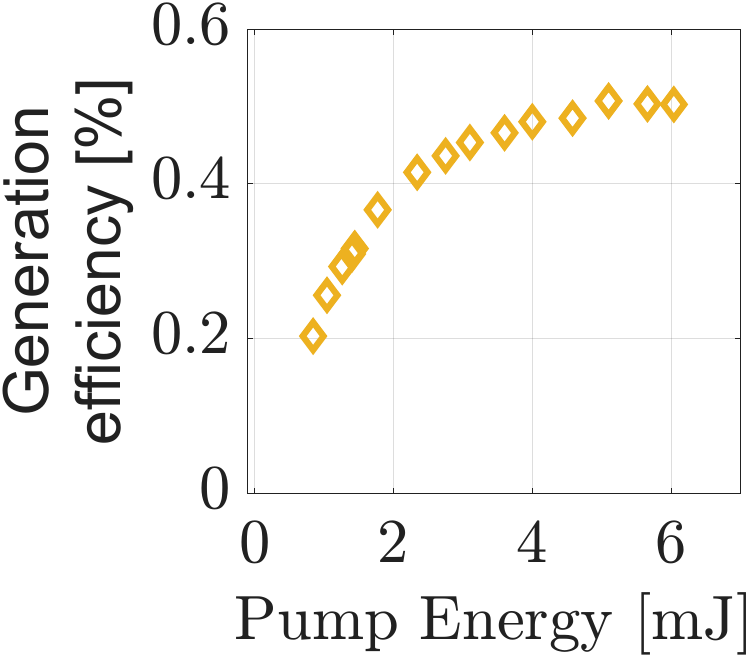}}
\end{picture}
\caption{Measured streaking gradient and THz pulse energy as a function of infrared pump energy. The inset shows the THz generation efficiency.}
\label{fig:streaking_vs_IR}
\end{figure}

Experimentally, the incident THz pulse energy is controlled using a motorized half-wave plate followed by a Brewster-angle window (see Fig.~\ref{fig:Experimental Setup}). The corresponding THz pulse energy is measured with a calibrated pyroelectric detector. Throughout these measurements, the electron beam is operated at the phase of maximum compression, corresponding to a kinetic energy of 4.6\,MeV. To place the streaking response on an absolute scale, full temporal scans are performed at selected pump energies, yielding streaking gradients of $21.3\,\mu\mathrm{rad/fs}$ at 1.7~mJ and $32.3~\mu\mathrm{rad/fs}$ at 4.7~mJ respectively. At other pump energies, the streaking gradient is obtained by scaling the measured beam deflection relative to these anchor measurements. The resulting trend, shown in Fig.~\ref{fig:streaking_vs_IR}, follows the expected square-root dependence on pump energy up to about $5\,\mathrm{mJ}$, consistent with the THz field amplitude scaling with the optical drive. The smaller than expected deflection observed at the highest pump energy is likely due to the limit in observable streaking strengths imposed by the waveguide geometry which are discussed below.

\begin{figure}[ht]
    \includegraphics[width = 0.95\columnwidth]{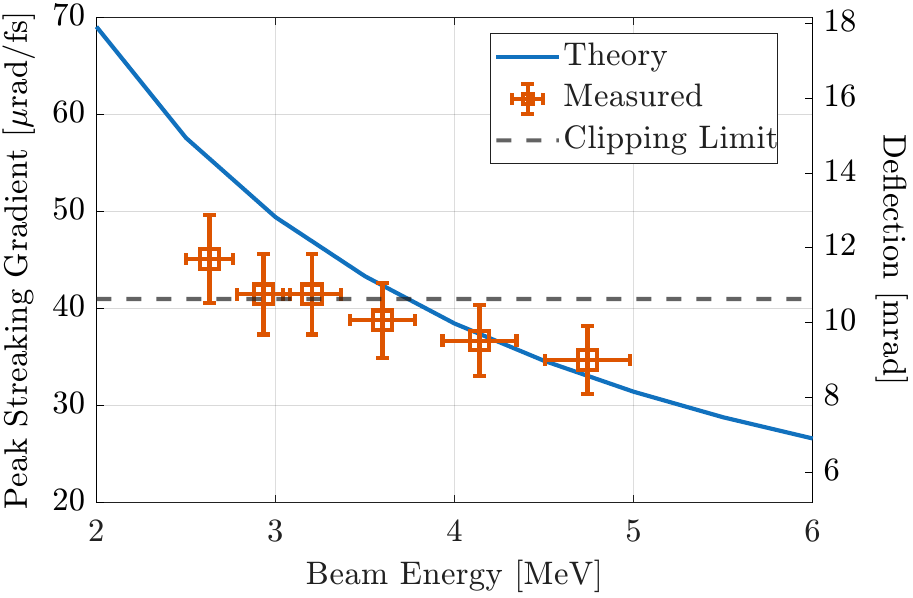}
    \caption{Measured deflection angle on the right and inferred streaking gradient on the left as a function of beam energy, demonstrating the expected inverse energy scaling.}
    \label{fig:streaking_vs_gamma}
\end{figure}

The beam-energy scaling predicted by Eq.~\eqref{eq:shear_strength} was examined by varying the linac phase. For beam kinetic energies above 4\,MeV, the streaking gradient was determined directly from differentiating the time-waveforms. At lower energies, the bunch length exceeds 500\,fs, making a full streaking calibration unreliable; in this regime, the transverse deflection was measured and the streaking gradient inferred assuming that the temporal profile of the deflecting waveform does not change with beam energy. As shown in Fig.~\ref{fig:streaking_vs_gamma}, the measured response follows the expected $1/E_b$ scaling within experimental uncertainty.

Although the inverse scaling would nominally imply streaking gradients approaching $60\,\mu\mathrm{rad/fs}$ at lower beam energies, the measurable gradient is ultimately constrained by the finite aperture of the structure. With a vertical waveguide gap of $50\,\mu\mathrm{m}$ over a 2.4\,mm interaction length, beam clipping becomes significant once the angular deflection reaches roughly $10\,\mu\mathrm{rad}$. As the streaking gradient increases, transmission through the structure correspondingly decreases, and for $S \gtrsim 39\,\mu\mathrm{rad/fs}$ the beam is transmitted only weakly near the extrema of the deflecting waveform. Beyond this point, increasing the THz pulse energy no longer produces a proportional increase in the measured streaking gradient. The observed saturation therefore arises from geometric transmission losses, not from an intrinsic limit of the streaking process.

The relation in Eq. \ref{eq:shear_strength} between the measured shearing strength and the transverse deflecting voltage allows to extract a beam-energy-independent figure of merit useful for comparing different streaking structures independent of beam energy. Using this procedure, we obtain \(V_\perp = (52\pm3)~\mathrm{kV}\) for the rectangular waveguide horn and \(V_\perp = (33\pm3)~\mathrm{kV}\) for the PPWG at 4.6~MeV in fair agreement with the predictions of Sec.~\ref{sec:theory}. Although these voltages are modest compared with those of RF transverse deflecting cavities, which routinely reach the multi-megavolt level \cite{Deflector:prat2025attosecond:polarix}, they provide a meaningful basis for quantitative comparison between different streaking approaches..

While $V_\perp$ is a useful structure-level figure of merit, temporal resolution is governed by the product $\omega V_\perp$. For the rectangular waveguide horn, the dominant spectral content lies near 0.5–0.6\,THz, yielding $\omega V_\perp \approx 0.18\,\mathrm{MV/ps}$. By comparison, the PolariX transverse deflecting structure at SwissFEL operates at 12\,GHz and achieves $\omega V_\perp \approx 5.5\,\mathrm{MV/ps}$, approximately a factor of 30 larger. However, PolariX employs an interaction length of approximately 2.4~m, three orders of magnitude longer than the present THz structure. When normalized to interaction length, the THz streaker delivers comparable shear per unit length while occupying a millimeter-scale footprint. This compactness, combined with intrinsic laser-to-beam synchronization and broadband single-cycle operation, represents a key advantage of THz-driven streaking. In addition, we should note that further optimization of geometry, confinement, and THz drive energy could substantially enhance performance while preserving the compact and broadband advantages of THz streaking.

\section{Resolution} 

\begin{figure}[ht]
     \includegraphics[width = 0.95\columnwidth]{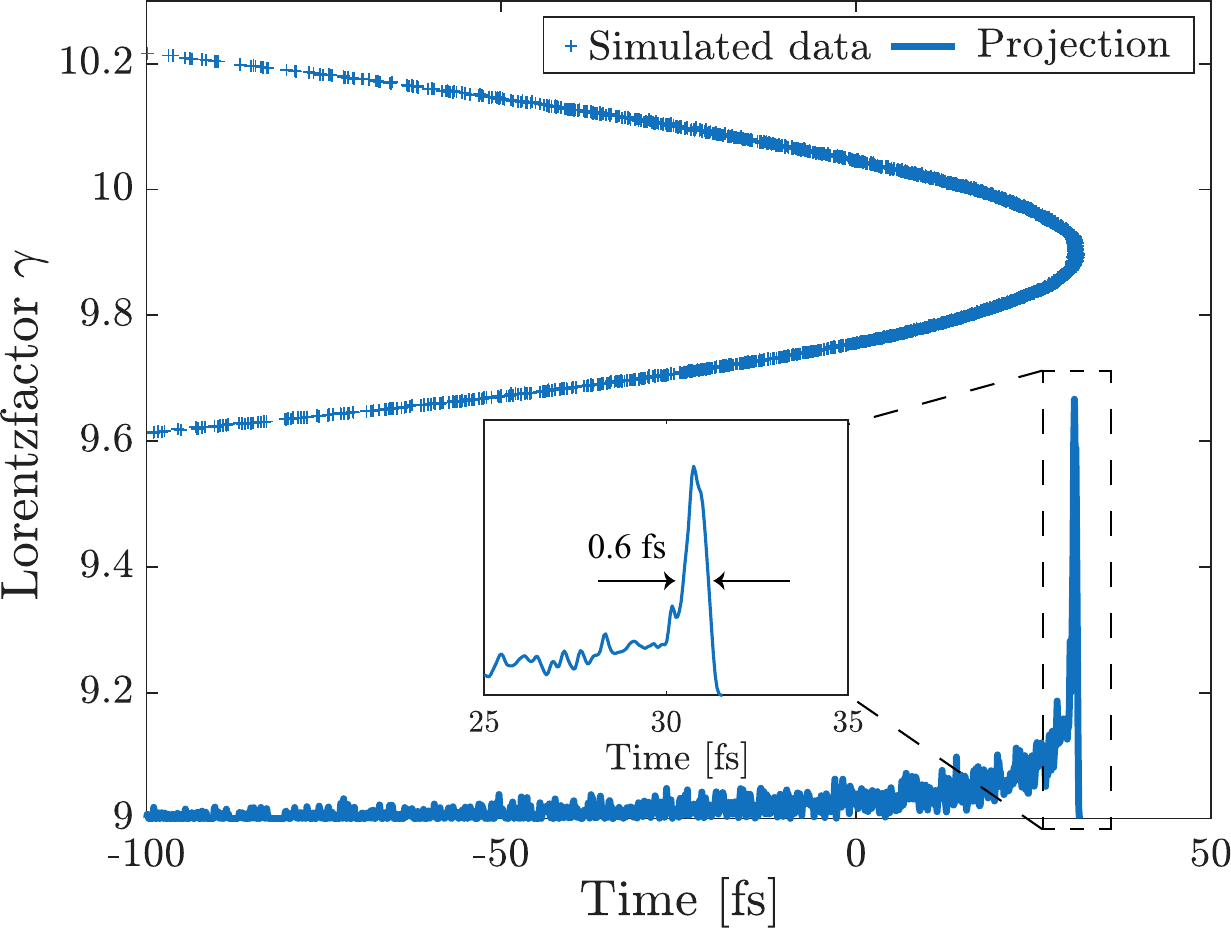}
     \caption{Simulated longitudinal phase space at the compression plane and
     corresponding temporal density projections illustrating sub-femtosecond
     density spikes.}
    \label{fig:GPT-sim}
\end{figure}

To probe the practical resolution of the THz streaking diagnostic, the beam is operated near maximum compression. Start-to-end GPT simulations predict that nonlinear compression of initially few-ps bunches produces strongly non-Gaussian current profiles with sub-femtosecond density spikes substantially shorter than the rms bunch length, even in the absence of additional linearization. The simulated longitudinal phase space and corresponding temporal projections at peak compression are shown in Fig.~\ref{fig:GPT-sim}. This distribution provides a very sharp feature which can be used to test the temporal resolution of the THz streaker. 

\begin{figure}[ht]
    \includegraphics[width = 0.95\columnwidth]{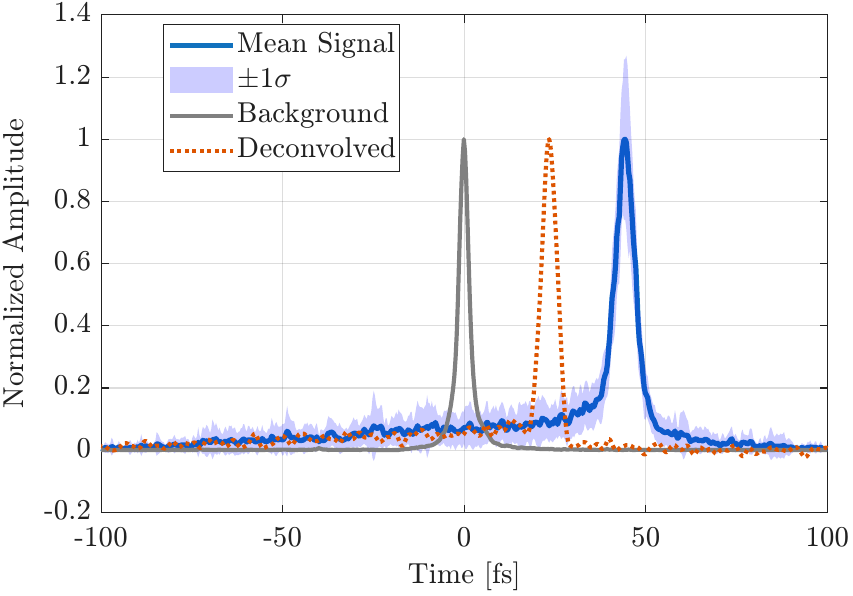}
    \caption{Averaged reconstructed temporal distributions for compressed bunches.
    Shaded region indicates shot-to-shot standard deviation (blue-solid). Unstreaked beam distribution (black-dashed) and Wiener-deconvolved bunch temporal profile (red-dotted)}
    \label{fig:deconvolved_signal}
\end{figure}

The background, unstreaked profile and the maximally streaked distributions for compressed bunches are shown in Fig.~\ref{fig:deconvolved_signal}. In order to account for the jitter in the compression, we collect data by bunch length (6–8\,fs), and then proceed to temporally realign, and average all the waveforms (31 shots) within each group to reduce statistical noise. The shaded region indicates the shot-to-shot standard deviation, providing a direct estimate of the statistical noise level. 

The temporal resolution of a transverse deflecting diagnostic is fundamentally limited by the convolution of the true temporal charge distribution with the unstreaked beam distribution and the detection noise. The streaked distribution observed on the screen, $n_y(t)$ (with $t = y/(L S)$ mapped from the transverse coordinate $y$ via the drift length $L$ and streaking gradient $S$), can be written as
\begin{equation}
n_y(t) = n_t(t) \ast n_0(t),
\end{equation}
where $n_t(t)$ is the true temporal density profile and $n_0(t)$ is the unstreaked beam distribution projected onto the streaking axis. Reconstruction of $n_t(t)$ therefore requires deconvolution.

\begin{figure}[ht]
    \includegraphics[width = 0.95\columnwidth]{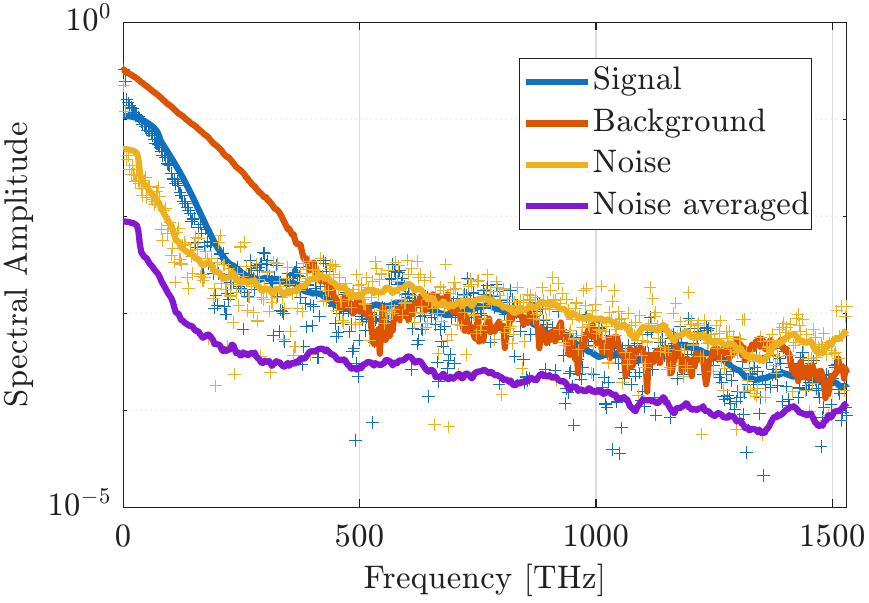}
    \caption{Fourier spectra of the unstreaked distribution, reconstructed signal,
    and noise floor derived from the compressed-beam dataset. The spectra for the signal and noise are smoothed to analyze crossings while the unsmoothed datapoints indicated as crosses.}
    \label{fig:noise-levels}
\end{figure}

To suppress the noise amplification inherent in direct spectral division, we employ Wiener deconvolution \cite{Signals:wiener1949extrapolation}. In the frequency domain, the reconstructed profile is given by
\begin{equation}
\tilde{n}_{t,\mathrm{est}}(\omega)=\frac{\tilde{n}_y(\omega)\tilde{n}_0^*(\omega)}{|\tilde{n}_0(\omega)|^2 + \mathcal{R}(\omega)},
\end{equation}
where $\mathcal{R}(\omega) = |\tilde{N}(\omega)|^2 / |\tilde{n}_t(\omega)|^2$ represents the spectral noise-to-signal ratio, estimated from the data shown in Fig.~\ref{fig:noise-levels}. Since the true spectrum $\tilde{n}_t(\omega)$ is unknown, it is approximated by the measured spectrum $\tilde{n}_y(\omega)$ to evaluate $\mathcal{R}(\omega)$. In this picture $\tilde{n}_0(\omega)$ denote the Fourier transform of the unstreaked distribution and $\tilde{N}(\omega)$ the spectral noise floor of the detection system. The usable bandwidth is restricted to frequencies below the cutoff $\omega_c$ defined by
\begin{equation}
|\tilde{n}_0(\omega_c)| = |\tilde{N}(\omega_c)|,
\end{equation}
which yields an approximate temporal resolution limit \cite{Signals:nyquist1928certain}
\begin{equation}
t_r \approx \frac{4\pi}{\omega_c}.
\label{eq:temporal-resolution}
\end{equation}
corresponding to the shortest resolvable temporal feature distinguishable in the reconstructed profile. The Wiener filter therefore provides a regularized reconstruction while naturally enforcing the experimentally accessible bandwidth.

The spectral content of the unstreaked distribution, the streaked signal, and the noise level derived from the same dataset are shown in Fig.~\ref{fig:noise-levels}. The background images recorded without THz field yield an rms screen width of $(4.53\pm0.18)$\,px, corresponding to a time-equivalent width of $(1.48\pm0.06)$\,fs. Shot-to-shot fluctuations of this width are small due to the 50\,$\mu$m waveguide aperture and do not significantly degrade temporal resolution.

For a single shot, the crossing between \(|\tilde{n}_0(\omega)|\) and the noise floor occurs near 400\,THz, corresponding through Eq.~\eqref{eq:temporal-resolution} to a temporal resolution of approximately 1.25\,fs. Averaging over multiple shot in presence of timing jitter requires being able to realign the peaks, but reduces the noise floor approximately as \(1/\sqrt{N}\), further extending the usable temporal bandwidth. The pixel limited resolution of the setup is 0.32\,fs. To improve beyond resolution limit, one could in principle increase the drift distance between deflecting structure and screen to increase the angular deflection per pixel without needing to increase the streaking power. 


The deconvolved beam distribution is also shown in  Fig.~\ref{fig:deconvolved_signal}. The original averaged distribution exhibits a FWHM width of 6.63\,fs, whereas the deconvolved profile narrows to 4.93\,fs, corresponding to a Gaussian rms width of 2.09\,fs. Although GPT simulations predict sub-femtosecond density spikes, with durations \(\lesssim 0.6\)\,fs, and the intrinsic diagnostic bandwidth should in principle support sub-femtosecond resolution, the experimentally reconstructed bunch duration remains larger. This difference is not unexpected. The simulated value represents an idealized limit that assumes accurate beam matching, nominal RF fields, and perfect alignment through the compression and diagnostic sections. In the experiment, several effects can broaden or wash out the shortest temporal features. These include residual off-axis RF fields, imperfect cancellation of transverse-longitudinal correlations, space-charge-induced distortion near the temporal focus, sensitivity to alignment through the deflecting structure, and possible transverse nonuniformity of the streaking field sampled by the beam.

\section{Conclusion}
We have presented a systematic experimental study of THz-driven transverse deflecting structures for ultrafast diagnostics of relativistic electron beams. Horn-coupled rectangular waveguides and parallel-plate waveguide structures were characterized in terms of streaking gradient, transverse deflecting voltage, and beam-energy scaling. The measured streaking gradients follow the predicted linear dependence on THz pulse energy and inverse dependence on beam energy, validating the theoretical framework. For the optimized rectangular waveguide horn, peak streaking gradients exceeding $30\,\mu\mathrm{rad/fs}$ and an effective transverse voltage of $(52\pm3)\,\mathrm{kV}$ were achieved within a millimeter-scale interaction length.

The use of THz frequencies provides substantial temporal shear in a compact structure, even though the absolute deflecting voltage remains below that of state-of-the-art RF transverse deflecting cavities. When evaluated in terms of $\omega V_\perp$, the THz streaker approaches the shear-per-length performance of meter-scale RF systems while providing intrinsic laser-to-beam synchronization. This combination of compactness, high operating frequency, and optical synchronization makes THz streaking attractive for femtosecond and potentially sub-femtosecond diagnostics, particularly in beamlines where conventional high power RF deflectors are impractical.

Application of Wiener deconvolution to compressed electron beams demonstrates femtosecond-level resolving capability and reveals non-Gaussian temporal substructure consistent with nonlinear compression dynamics. A spectral-cutoff analysis gives an instrumental single-shot temporal resolution of approximately 1.3 fs for a single-shot measurement. Our results define the current performance limits of fast-wave THz streaking structures and establish practical design guidelines for further optimization. Promising pathways toward higher streaking gradients to achieve sub-femtosecond resolution also include emerging single-shot transverse interaction geometries \cite{THz:Streaking:zhao2019terahertz:thz-oscilloscope} and dispersion-free, dually tapered parallel-plate waveguides capable of supporting GV/m fields at the 10~$\mu$J level \cite{THz:Structure:iwaszczuk2012terahertz:dual-PPWG,THz:Structure:rohrbach2023wideband:dispersion-free,THz:Streaker:he2025terahertz:perpendicular}.

\begin{acknowledgments}
M.L, B.S., A.K. and P.M. have been supported by U.S. Department of Energy Grant No. DE-AC02-76SF00515. 
\end{acknowledgments}

\bibliography{bibliography}
\end{document}